\documentstyle[12pt,epsfig]{article}
\include{epsfig,wrapfig,times,mathptm}
\renewcommand{\section}[1]{\vspace{6pt} \noindent\mbox{#1} \newline \noindent}
\renewcommand{\subsection}[1]{\vspace{6pt} \noindent\mbox{\underline{#1}} 
\newline \noindent}
\renewcommand{\subsubsection}[1]{\vspace{6pt} \noindent\mbox{\underline{#1}}
\noindent}

\newfont{\sansb}{cmssbx10}
\newfont{\sans}{cmss10}

\begin{document}

{\center \ {\large OBSERVATION OF THE CRAB NEBULA GAMMA-RAY EMISSION ABOVE 220 GeV 
BY THE CAT CHERENKOV IMAGING TELESCOPE}
\vspace{30pt}\\

A.~Barrau$^5$, R.~Bazer-Bachi$^2$, H.~Cabot$^7$, L.M.~Chounet$^4$, G.~Debiais$^7$,
B.~Degrange$^4$, J.P.~Dezalay$^2$, A.~Djannati-Ata\"{\i}$^5$, 
D.~Dumora$^1$, P.~Espigat$^3$, B.~Fabre$^7$, P.~Fleury$^4$, G.~Fontaine$^4$, R.~George$^5$,
C.~Ghesqui\`ere$^3$, P.~Goret$^6$, C.~Gouiffes$^6$, I.A.~Grenier$^{6,9}$, L.~Iacoucci$^4$,
S.~Le Bohec$^3$, I.~Malet$^2$, C.~Meynadier$^7$, F.~Munz$^8$, T.A.~Palfrey$^{10}$, E.~Par\'e$^4$, 
Y.~Pons$^5$, M.~Punch$^3$, J.~Qu\'ebert$^1$, K.~Ragan$^1$, C.~Renault$^{6,9}$, 
M.~Rivoal$^5$, L.~Rob$^8$, P.~Schovanek$^{11}$, D.~Smith$^1$,  J.P.~Tavernet$^5$
and J.~Vrana$^{4 \dag}$ \vspace{6pt}\\
{\it $^1$Centre d'Etudes Nucl\'eaire de Bordeaux-Gradignan$^{\ast}$, France \\
$^2$Centre d'Etudes Spatiales des Rayonnements$^{\ddagger}$, Toulouse, France \\ 
$^3$Laboratoire de Physique Corpusculaire$^{\ast}$, Coll\`ege de France, Paris, France \\
$^4$Laboratoire de Physique Nucl\'eaire de Haute Energie$^{\ast}$, Ecole Polytechnique, Palaiseau, France \\
$^5$Laboratoire de Physique Nucl\'eaire de Haute Energie$^{\ast}$, Universit\'es de Paris VI et VII, France \\
$^6$Service d'Astrophysique$^{\#}$, Centre d'Etudes de Saclay, France \\
$^7$Groupe de Physique Fondamentale$^{\ast}$, Universit\'e de Perpignan, France \\
$^8$Nuclear Center, Charles University, Prague, Czech Republic \\
$^9$Universit\'e Paris VII \\
$^{10}$Department of Physics, Purdue University, Lafayette, IN 47907,
U.S.A \\
$^{11}$JLO Ac. Sci. \& Palacky University, Olomouc, Czech Republic \\
$^{\ast}$IN2P3/CNRS \\
$^{\ddagger}$INSU/CNRS \\
$^{\#}$DAPNIA/CEA \\
$^{\dag}$Deceased \\
}}

 {\vspace{12pt}}
{\center ABSTRACT\\}

The CAT imaging telescope, recently built on the site of the former solar plant
``Th\'emis'' (French Pyr\'en\'ees), observed 
$\gamma$-rays from the Crab nebula from October 1996 to March 1997. 
This steady source, often considered as the standard candle of very-high-energy
$\gamma$-ray astronomy, is used as a test-beam to probe the performances of the new 
telescope, particularly its energy threshold ($220 \: GeV$ at $20^\circ$ zenith angle) 
and the stability 
of its response. Due to the fine-grain camera, an accurate 
analysis of the longitudinal profiles of shower images is performed, yielding the source
position in two dimensions for each individual shower.

\vspace{6pt}
\setlength{\parindent}{1cm}
\section{INTRODUCTION}

 With a reflector of only
$ 17.7 \: m^2 $, the CAT imaging telescope
was designed to achieve a rather low threshold ($ \sim 220 \: GeV$)
by taking full advantage of the rapidity of the Cherenkov signal with an almost 
isochronous mirror, fast phototubes and fast trigger and readout electronics. Moreover,
its very-high-definition camera (546 pixels with an angular size of $0.12^\circ$) allows 
a powerful image analysis (Degrange and Le Bohec 1995) leading to a high rejection of hadron showers. 
A further advantage is that the source position on the sky can be reconstructed for each shower 
on the basis of the longitudinal distribution of light in its image, the accuracy being close
to the pixel size. The position of a point source can then be determined to about $ 1 \: arc \: min $
with about $100 \: \gamma$-ray showers (Le Bohec 1996).

The Crab nebula is the best-studied source of very-high-energy $\gamma$-rays (Lewis et al. 1993, 
Djannati-Ata\"{\i}
1995, Deckers et
al. 1995) and its emission is found to be steady within the sensitivity of
present instruments. It can thus be used as a test-beam to check the stability of the detector 
response and to characterize its energy threshold. Furthermore, it is important to validate the
simulations of all instrumental effects (reflector optics, light catchers, phototubes and electronics)
which are needed to determine the equivalent detection area as a function of primary energy 
and zenith angle, as well as the accuracy in energy measurement, all quantities necessary to measure
the energy spectrum of the source. Such a validation is provided by comparing $\gamma$-ray event rates
from the Crab nebula obtained experimentally with the corresponding quantities predicted by 
simulations on the
basis of the energy
spectrum given by
previous
experiments, (Lewis
et al. 1993,
Djannati-Ata\"{\i} 1995).

\vspace{6pt}
\section{TRIGGER CONDITIONS}

The CAT triggering system uses a majority logic based on the 288 inner phototubes of the camera.
The pulse heights of at least $4$ phototubes were required to be
greater than the average level of a $3$-photoelectron signal, in at least one of 9 mutually 
overlapping sectors of the camera, each one corresponding to about $60^\circ$.
With such conditions, random coincidences 
due to the sky noise are practically eliminated and the trigger rate, mainly due to hadronic showers and
single muons, is $\sim \: 15 \: Hz$ at moderate zenith angles ($20^\circ$ to $ 30^\circ$). These conditions
do not correspond to the minimal threshold of the telescope, since the
detector can be operated with trigger rates up to $\sim \: 50 \: Hz$.

\vspace{6pt}
\section{EVENT SELECTION AND BACKGROUND SUBTRACTION}

Data have been submitted to two different image analyses :

\subsubsection{Analysis based on the first and second moments of the light distribution}

The first method, based on the first and second moments of the light distribution in the image, has been
successfully used by
the Whipple group
(Punch et al. 1991). Each shower image is characterized by its angular ``length''
$L$, ``width'' $W$, ``distance'' $D$ (i.e. distance from the centroid of the image to the center of the field of 
view) and by its
``size'' $S$
(i.e. the total
integrated light
content of the
shower) (Fegan 1991). The cuts in 
these variables have been roughly adapted to the resolution offered by the CAT camera, but not optimized.
The background due to hadronic
showers as well as to single muons is severely reduced by the following requirements : $ 2 < L < 5$ ;
$ 0.7 < W < 1.5$ ; $D > 8.5$ (in which angles are expressed in milliradians)
and $S > 30$ photoelectrons. Then, the  
distribution of the orientation angle $ \alpha $ (i.e. the angular deviation between the image principal 
direction and the line connecting the centroid of the image to the center of the field of view) shows a clear
$\gamma$-ray signal at low values of $\alpha$ ($ \alpha \: < \: 9^\circ$) above an almost flat background, 
which can be subtracted by using off-source data (figure 1a).

\subsubsection{Analysis based on a maximum-likelihood method}

The second analysis consists of a maximum-likelihood method (Degrange and Le Bohec 1995) (Le Bohec 1996), based on an analytical
model of
$\gamma$-ray showers
proposed by Hillas
(Hillas 1982). For a genuine $\gamma$-ray shower, this model
provides the average distribution of Cherenkov light in the focal plane as a function of several parameters : the
azimuth angle $\phi$ of the image axis, the $\gamma$-ray energy $E_{\gamma}$, the distance $R$ between the 
shower axis and the telescope (or impact parameter), and finally the two angular coordinates of the source. 
In this method, the source position is thus left as a free parameter which
can be reconstructed on a shower per shower basis, due to its influence on the longitudinal light distribution. 
The analytical model is used to define a $\chi^2$-like function of $\phi$, $E_{\gamma}$, 
$R$ and source position. Fluctuations in the pixel signals are estimated from Monte-Carlo studies of 
$\gamma$-ray-induced showers by requiring that the $\chi^{2}$ probability distribution be approximately uniform.
For each shower observed, this function is minimized with respect to
the parameters: the fitted $\chi^{2}$ provides gamma-hadron discrimination; the $\gamma$-ray 
energy determination automatically takes account of the shower position with respect to the 
telescope which is obtained from the same fit, which further yields the source position in 
$2$ dimensions. For those events with a $\chi^{2}$ probability greater than $0.35$, the distribution of the 
orientation angle 
$\alpha$ is shown in figure 1b. Since the method discriminates between the top and the bottom of showers,
this angle is here plotted from $0^\circ$ to $180^\circ$. The $\gamma$-ray signal clearly shows up at low values 
of $\alpha$ ($ \alpha \: < \: 9^\circ$) whereas the almost flat hadronic or muonic background can easily be 
subtracted by using off-source data. The 2-dimensional distribution of the reconstructed source
position is shown in figure 2 in celestial coordinates centered on the Crab nebula, after background subtraction,
showing that a single imaging telescope with a very-high-definition camera has a real source localization capability.
\vspace{6pt}

\section{GAMMA-RAY EVENT RATES}

The simulation is based on a two-step procedure, one concerning the development of electromagnetic air showers
(Kertzman and
Sembroski 1994) and the other the detection of Cherenkov photons and subsequent instrumental effects (Le Bohec 1996). In the present
study whose goal is
to validate our
knowledge of the
detector response,
the Crab nebula
spectrum quoted in Lewis et al. 1993 was used as input to simulate $\gamma$-ray-induced
showers above $50 \: GeV$ with an integral spectral index of 1.7.
Losses in $\gamma$-ray events due to
the cuts described above can easily be modeled in the moment-based analysis, whereas the second method requires a more
accurate treatment of the night-sky background; we thus restrict ourselves here to the first method. For
those observations with a zenith angle smaller than $30^\circ$, the raw $\gamma$-ray rate (i.e. after cuts) from the
Crab Nebula is found to be $2.0 \pm 0.3 \: \gamma$-ray per minute. Correcting for losses due to cuts, the total rate
is found to be $5.1 \pm 0.7 \: \gamma$-ray per minute.
This corrected rate characterizes the telescope response 
and particularly its energy threshold. Since the definition of the threshold of an Atmospheric Cherenkov Telescope 
varies among
authors, the corrected $\gamma$-ray event rates from a steady source provide
an objective way of comparing different instruments. The Monte-Carlo expectations for the corrected event 
rates ($6.4\: \gamma$-ray per minute at $20^\circ$ zenith angle and $4.7\: \gamma$-ray per minute at $30^\circ$)
are found in good agreement with the experimental value, showing that
the instrumental simulation provides a reasonable description of the apparatus. If the threshold is
defined as the energy of maximal $\gamma$-ray-event rate from the Crab nebula, the corresponding value for CAT with
the triggering conditions quoted above is $220 \: GeV$ at $20^\circ$ zenith angle. 
Finally, figure 3
shows that $\gamma$-ray event rates obtained in similar conditions of zenith angle but at different times are
compatible with each other within statistical errors; the detector response is thus found to be stable, a necessary condition for the study of very variable sources such as the Active Galactic Nuclei Markarian 421 and
Markarian 501.
\vspace{6pt}

\section{CONCLUSION}

From the preceding study, we conclude that the CAT telescope, which started operation in October 1996,
achieves a
low energy threshold despite its moderate 
size, as shown by the
$\gamma$-ray event rates from the Crab nebula. Its response is
found to be stable in time. Furthermore, its very-high-definition camera provides a real source localization capability,
which will be essential in the study of poorly localized sources.
\vspace{6pt}

\newpage

\section{REFERENCES}
\setlength{\parindent}{-5mm}
\begin{list}{}{\topsep
0pt \partopsep 0pt \itemsep 0pt 
\leftmargin 5mm
\parsep 0pt
\itemindent -5mm} 
\vspace{-15pt}

\item Degrange, B. and Le Bohec, S., Proc. of the XXIV ICRC,
, p 436, Rome (1995).
\item Le Bohec, S., Ph.D. thesis, Universit\'e de Paris XI Orsay (1996).
\item Lewis, D.A., Akerlof, C.W., Fegan, D.J. et al., Proc. of the XXIII ICRC,
1, p 279, Calgary (1993).
\item Djannati-Ata\"{\i}, A., Proc. of the XXIV ICRC,
2, p 315, Rome (1995).
\item Deckers, T., Proc. of the XXIV ICRC,
2, p 319, Rome (1995).
\item Punch, M., Akerlof, C.W.,
Cawley, M.F., et al., Proc. of the XXII ICRC,				
1, p 464, Dublin (1991).
\item Fegan, D.J., Proc. of the First Workshop ``Towards a Major Atmospheric Cherenkov
Detector'', p 3, Palaiseau (1991), edited by P. Fleury and G. Vacanti.
\item Hillas, A.M., J. Phys. G, Nucl. Phys., 8, 1461 (1982).
\item Kertzman, M.P. and Sembroski, G., Nucl. Inst. and Meth., A343, 629 (1994).
\end{list}

\vspace{5cm}
\begin{figure}[htb]
\centering
\mbox{\epsfig{figure=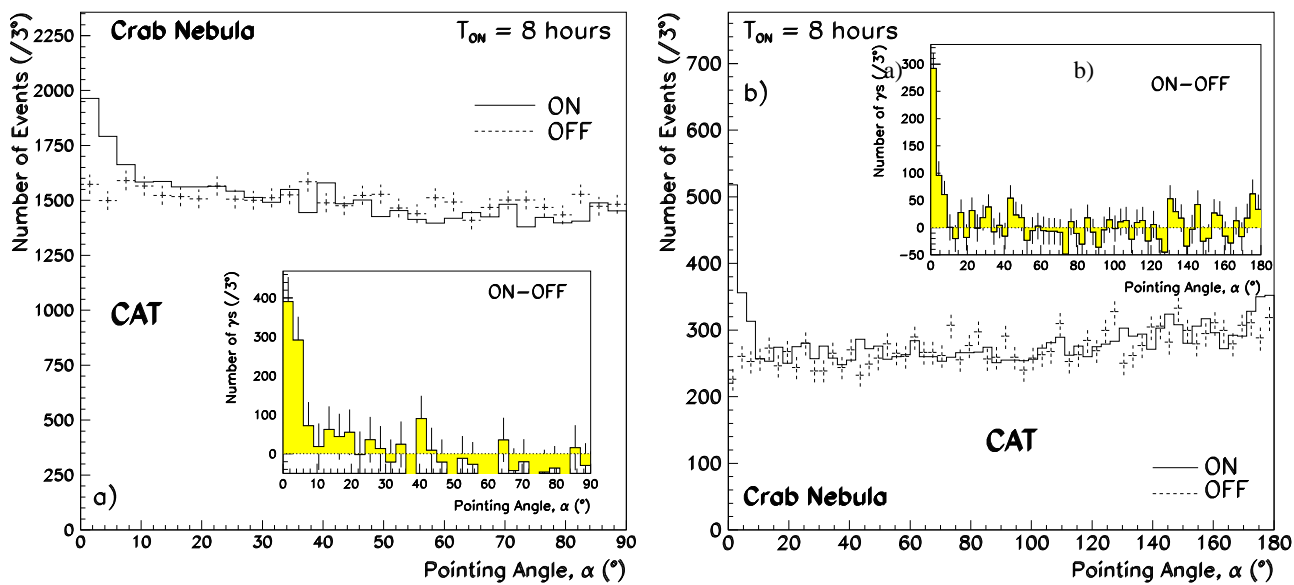}}
%,bbllx=1.35cm,bblly=5.4cm,bburx=12.75cm,bbury=12.9cm}
\caption[Orientation angle $\alpha$ for 8 hours of Crabe observation
: (a) moment based analysis and (b) maximum-liklihood method.  ]
{Orientation angle $\alpha$ for 8 hours of Crab observation
: (a) moment based analysis and (b) maximum-likelihood method. }
\label{101}
\end{figure}                                             

\begin{figure}[htb]
\centering
\mbox{\epsfig{figure=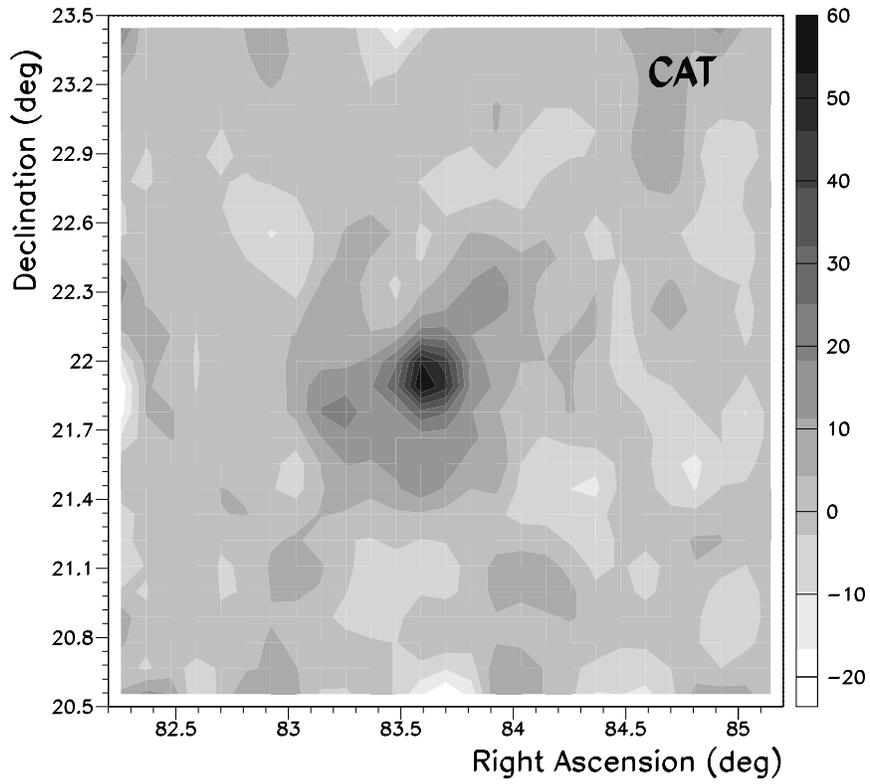,height=11.5cm}}
%,bbllx=1.35cm,bblly=5.4cm,bburx=12.75cm,bbury=12.9cm}
\caption[Reconstructed source position after background subtraction
in celestial coordinates for the Crab nebula ]
{Reconstructed source position after background subtraction
in celestial coordinates for the Crab nebula. }
\label{102}
\end{figure}                                             

\begin{figure}[htb]
\centering
\mbox{\epsfig{figure=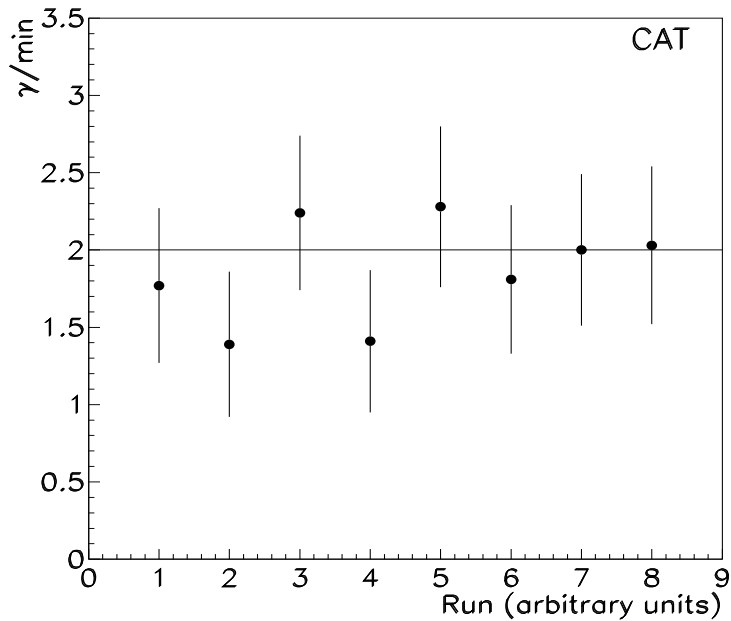,height=9cm,width=10cm}}
%,bbllx=1.35cm,bblly=5.4cm,bburx=12.75cm,bbury=12.9cm}
\caption[$\gamma$-ray
event rates for $90$
minute periods in similar
conditions of zenith angle ($20^\circ$ - $30^\circ$, December 1996 to
Febuary
1997). Horizontal
scale is arbitrary.]
{ $\gamma$-ray
event rates for $90$
minute periods in similar
conditions of zenith angle ($20^\circ$ - $30^\circ$, December 1996 to
Febuary
1997). Horizontal
scale is arbitrary.}
\label{103}
\end{figure}

\end{document}